\documentclass{article}
\usepackage{epsf,amsmath}
\usepackage{graphicx}
\usepackage{epsfig}
\usepackage{color}
\usepackage{cite}

\begin{document}

\title{Small-x QCD physics probed with jets in CMS
\thanks{Presented at  the Low x workshop, May 30 - June 4 2013, Rehovot and
Eilat, Israel}%
% you can use '\\' to break lines
}
\author{Pedro Cipriano\footnote{pedro.cipriano@desy.de}, on behalf of the CMS Collaboration
\\
%\address
{\small
Deutsches Elektronen-Synchrotron, Notkestrasse 85, 22607 Hamburg, Germany
}
\smallskip\\
}
\date{\today
}
\maketitle

\maketitle

\begin{abstract}
The latest CMS jet measurements in p-p collisions at $\sqrt{s}$ = 7 TeV, sensitive 
to small-x QCD physics, are discussed.
These include inclusive forward jet and simultaneous forward-central jet production, as
well as production ratios and azimuthal angle decorrelations of jets widely separated in rapidity. 
\end{abstract}

%\markright{}

%\renewcommand{\@evenhead}

\markboth{\large \sl \hspace*{0.25cm}\underline{C. Royon} 
\hspace*{0.25cm} } {\large \sl \hspace*{0.25cm} Low-x 2013 Proceedings}

\section{Introduction}
\begin{figure}[ht]
\begin{center}
\includegraphics[height=0.4\textheight]{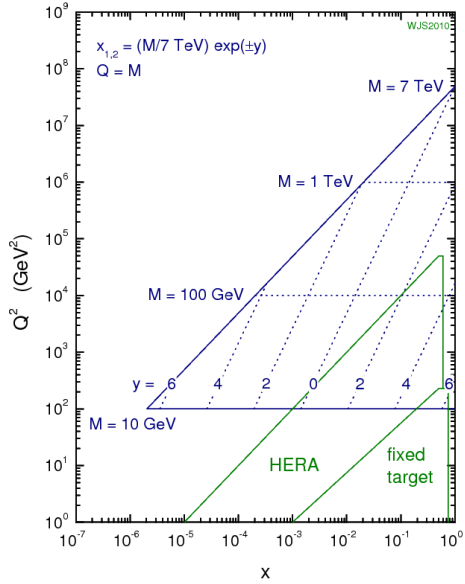}
\caption{Kinematic phase--space accessible to Hera and LHC \cite{parton_fig}.}
\label{fig:intro}
\end{center}
\end{figure}

The measurement of forward jets provides an 
important testing ground for QCD predictions of the Standard Model in the low-x region. The 
LHC (Large Hadron Collider) can reach $Q^2$ and 
$x$ values previously inaccessible to Hera as displayed in figure \ref{fig:intro}. To access 
the low-x region one must
look at high rapidity. For such task the rapidity coverage of up to 
$|\eta| $ = 5.2 in CMS \cite{cms} has been used.

The jet--rapidity and transverse--momenta is well described by the calculations at next-to-
leading-order (NLO) in perturbative quantum chromodynamics (QCD) using the 
Dokshitzer-Gribov-Lipatov-Altarelli-Parisi (DGLAP) 
\cite{dglap1, dglap2, dglap3, dglap4, dglap5} approach and collinear factorization.
The dijet cross-section is also well described \cite{dijets_2011}. When the collision energy $\sqrt{s}$ is considerably 
larger than the hard scattering  scale given by the
jet transverse momentum, $p_{T}$,
calculations in perturbative QCD require a resummation of large $\log(1/x)$ terms. This leads
to the prediction of new dynamic effects, expected to be described by Balitsky-Fadin-Kuraev-Lipatov 
(BFKL) evolution \cite{bfkl1, bfkl2, bfkl3} and $k_{T}$ factorization \cite{ktfact1, ktfact2, ktfact3}.
An effective theory has been developed which describes strong interactions in this kinematic domain \cite{smallx}.
This description is particularly useful in events with several jets with large rapidity separation, 
which are not well described by DGLAP predictions.

To extend the study of the parton evolution equations, the azimuthal angle 
differences were also measured. This observable
has a sensitivity to BFKL effects when both jets are widely separated in rapidity (eg: Mueller-Navelet jets).

\section{Inclusive forward jet production}

The inclusive forward jet cross-section was measured from an integrated luminosity of 3.14 $pb^{-1}$~\cite{jets1}. Jets 
were reconstructed 
with the anti-$k_{T}$ clustering algorithm \cite{antikt,fastjet} with a distance parameter R = 
$\sqrt{(\Delta\eta)^2 + (\Delta\phi)^2}$ = 0.5. The energy depositions in the calorimeter cells were used as input for the 
clustering. Assuming massless jets, a four--momentum is associated with them by summing the energy of the cells above a
given threshold.

\begin{figure}[htp]
\begin{center}
\includegraphics[width=0.6\textwidth]{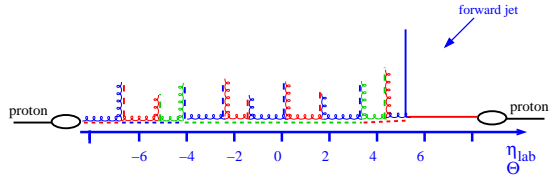}
\caption{Feynman diagram for inclusive forward jet production}
\label{fig:feynman_inclusive_forward}
\end{center}
\end{figure}

The forward region is defined as 3.2 $ < |\eta| < $ 4.7. The jets are required to have a transverse momentum above $p_{T}$ = 35 GeV.
If more than one jet is present, the one with with highest $p_{T}$ is considered, as is illustrated in figure 
\ref{fig:feynman_inclusive_forward}. The jets are corrected for the following 
systematic effects: $p_T$ and $\eta$--dependent response of the calorimeters, overlap with other proton--proton interactions
and the migration of events across the $p_{T}$ bins due to jet energy resolution.

\begin{figure}[htp]
\begin{center}
\includegraphics[width=0.36\textwidth]{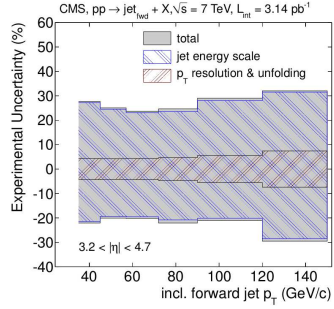}
\caption{Inclusive forward jet production uncertainty~\cite{jets1}.}
\label{fig:inclusive_forward_unc}
\end{center}
\end{figure}

In figure \ref{fig:inclusive_forward_unc} the experimental systematic uncertainties are shown for the leading forward jet 
as function of $p_T$. The jet energy 
scale is the dominant systematic uncertainty and the total uncertainty is around -25+30\%.

\begin{figure}[htp]
\begin{center}
\includegraphics[width=0.85\textwidth]{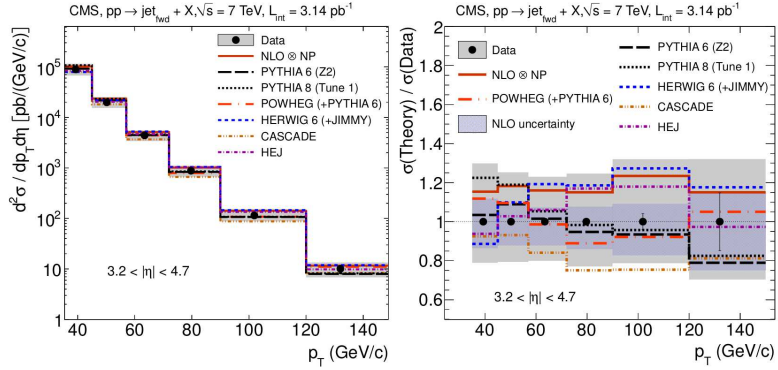}
\caption{Inclusive forward jet cross-section compared with different Monte Carlo predictions~\cite{jets1}.}
\label{fig:inclusive_forward}
\end{center}
\end{figure}

The inclusive forward jet production cross--section corrected to hadron level is presented in 
figure \ref{fig:inclusive_forward}.
Although all predictions describe the data within the uncertainty band, some of them do 
better. {\sc Powheg}~\cite{powheg}~+ {\sc Pythia~6}~\cite{pythia6}
gives the best description. {\sc Pythia 6} and {\sc Pythia 8}~\cite{pythia8} describe the data reasonably 
well. {\sc Cascade}~\cite{cascade} underestimates the cross-section while 
{\sc Herwig 6}~\cite{herwig} + {\sc Jimmy}~\cite{jimmy} tends to overestimate. NLOJET++ overestimates the data but is still within the 
large theoretical and experimental uncertainties.

\section{Forward-central dijet production}

\begin{figure}[htp]
\begin{center}
\includegraphics[width=0.6\textwidth]{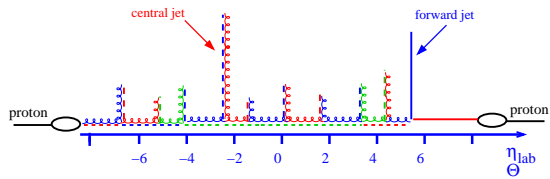}
\caption{Feynmann diagram for forward--central dijet production}
\label{fig:feymamn_forwardcentral}
\end{center}
\end{figure}

The selection procedure for the simultaneous forward--central dijet production is similar to the one for for the 
inclusive forward jet 
production. In addition, a central jet within $|\eta| <$ 2.8 with  a transverse 
momentum above $p_T$ = 35 GeV is required. A Feynman diagram
of the process is shown in figure \ref{fig:feymamn_forwardcentral}.

\begin{figure}[htp]
\begin{center}
\includegraphics[width=0.8\textwidth]{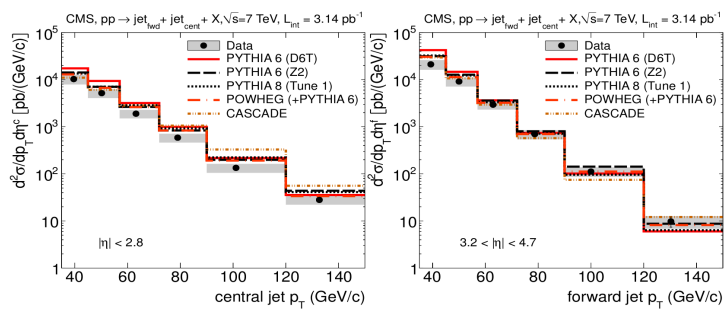}
\caption{Forward--central dijet production compared with different Monte Carlo predictions \cite{jets1}.}
\label{fig:forwardcentral1}
\end{center}
\end{figure}

\begin{figure}[htp]
\begin{center}
\includegraphics[width=0.8\textwidth]{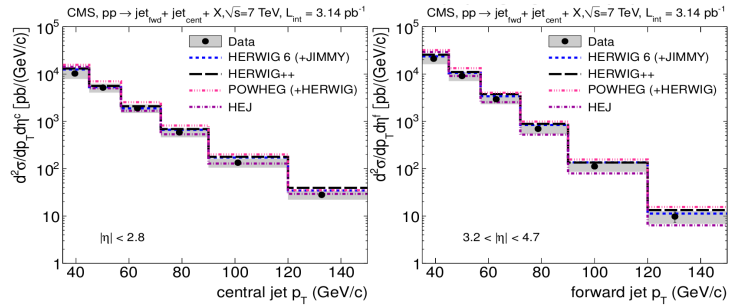}
\caption{Forward--central dijet production compared with different Monte Carlo predictions \cite{jets1}.}
\label{fig:forwardcentral2}
\end{center}
\end{figure}

Several MC predictions compared to the data cross-section is presented in 
figures \ref{fig:forwardcentral1} and \ref{fig:forwardcentral2}~\cite{jets1}. Forward jet cross-section is steeper 
than the central jet. The shape of the forward jet is poorly described when compared with the central jet.
HEJ~\cite{hej} provides the best description being followed closely by {\sc Herwig 6} and
{\sc Herwig ++}~\cite{herwigpp}. Both {\sc Pythia 6}, {\sc Pythia 8} and the CCFM {\sc Cascade} have 
troubles describing the data for the central jets and for low $p_T$ forward jets. {\sc Powheg + Pythia 6}, 
which was the best prediction for 
inclusive forward jet production, yelds similar result as {\sc Pythia 6} alone.

\section{Azimuthal--angle decorrelations of jets widely separated in rapidity}

\begin{figure}[htp]
\begin{center}
\includegraphics[width=0.45\textwidth]{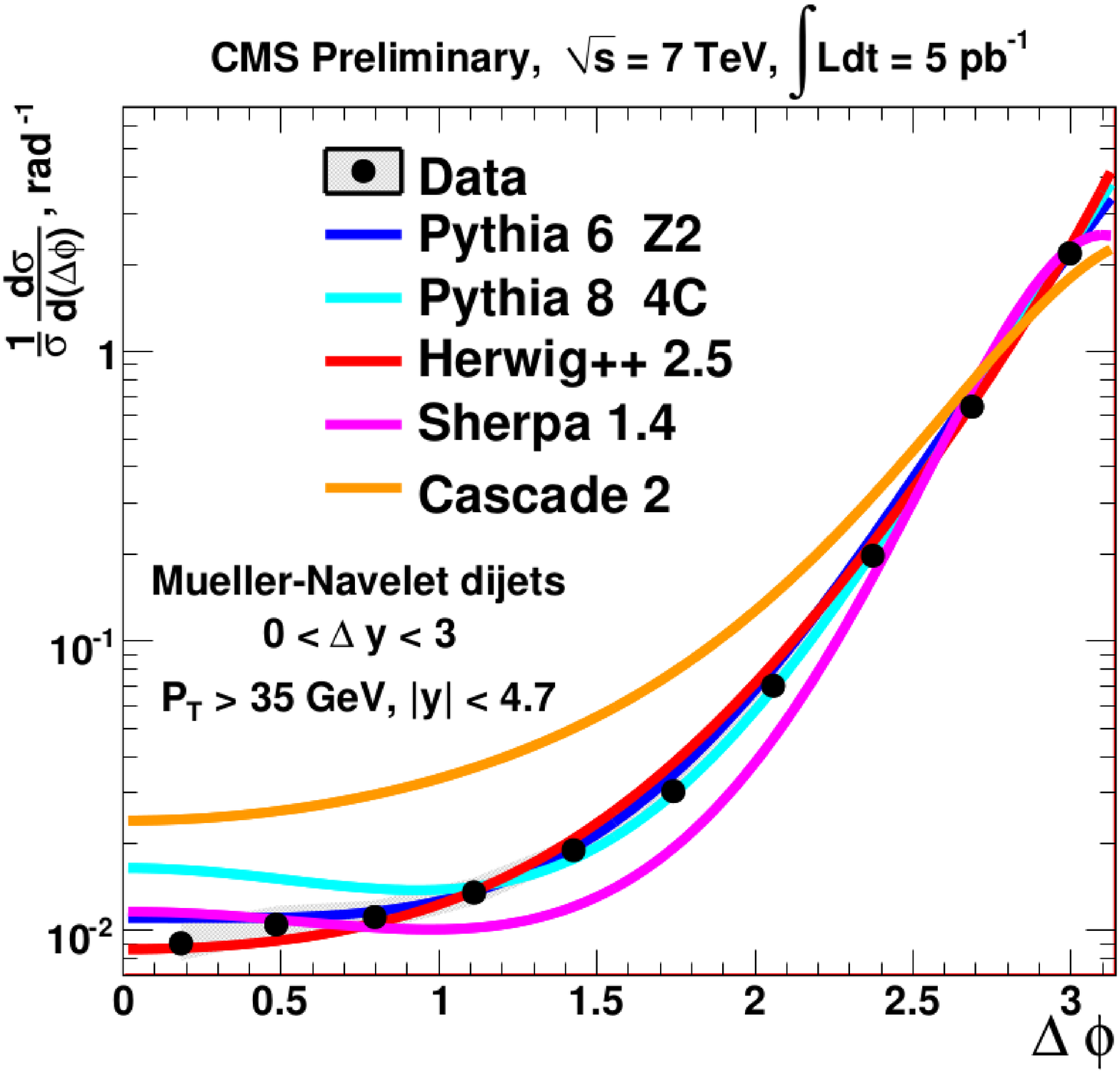}
\includegraphics[width=0.45\textwidth]{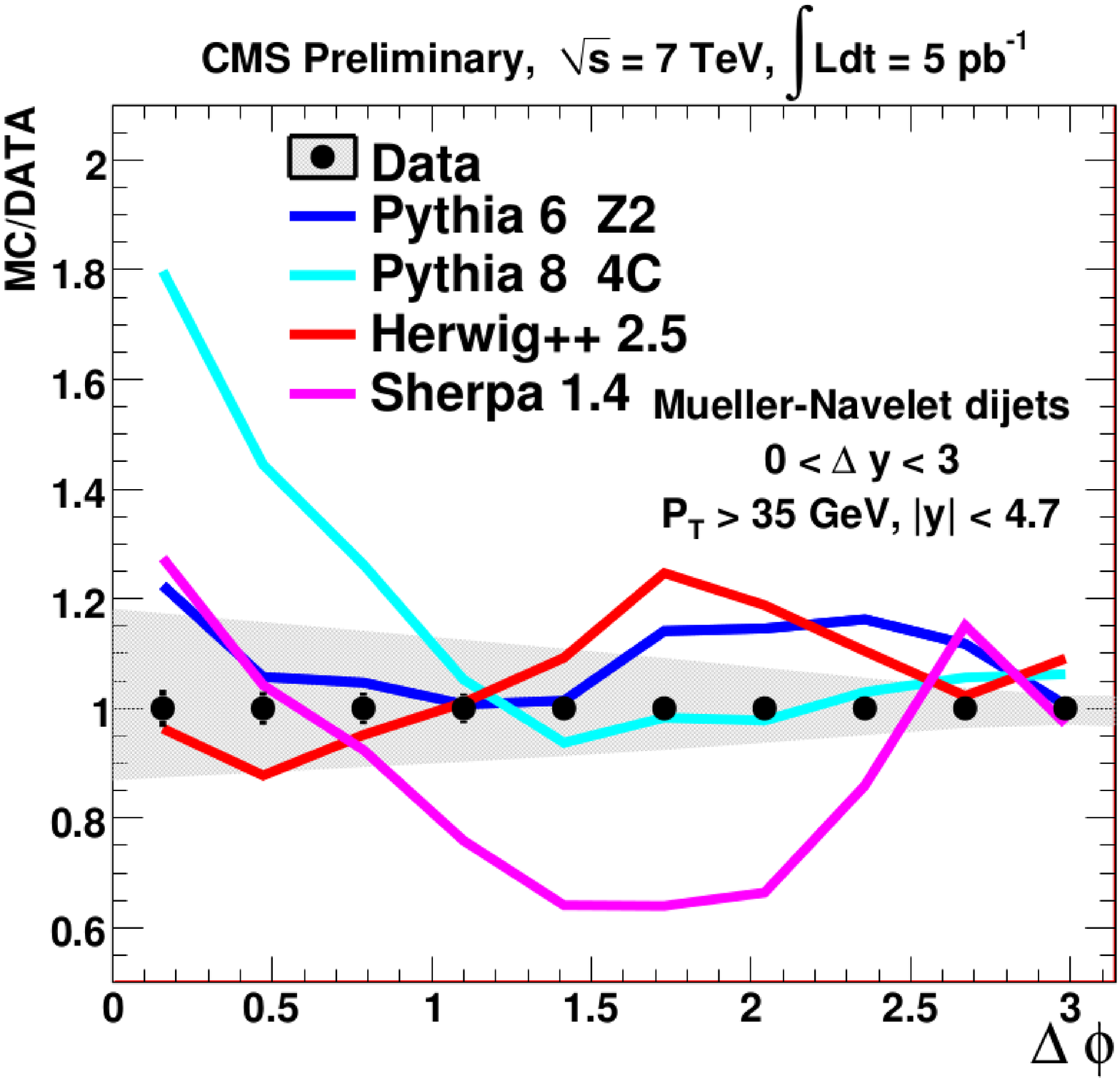}

\includegraphics[width=0.45\textwidth]{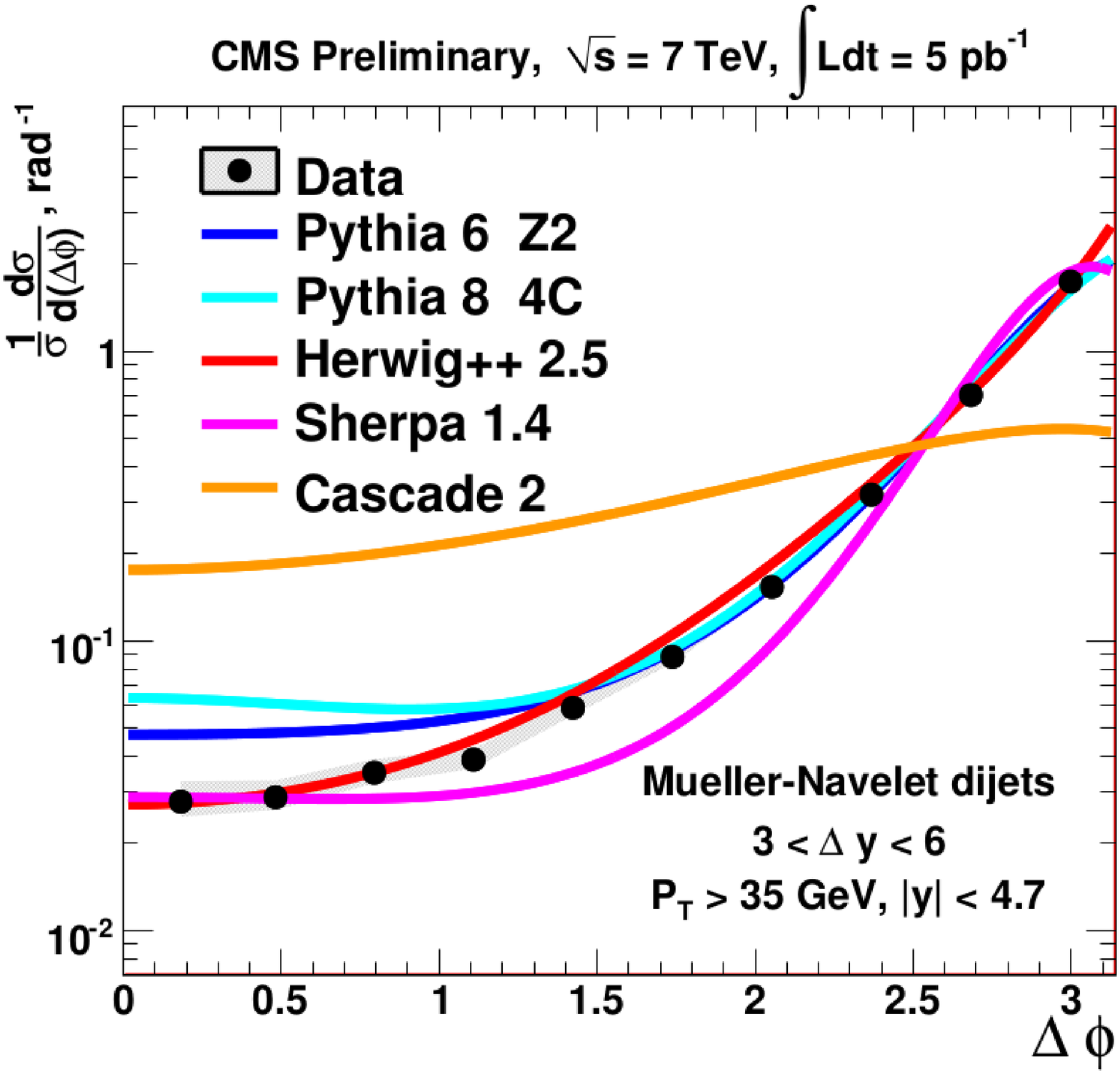}
\includegraphics[width=0.45\textwidth]{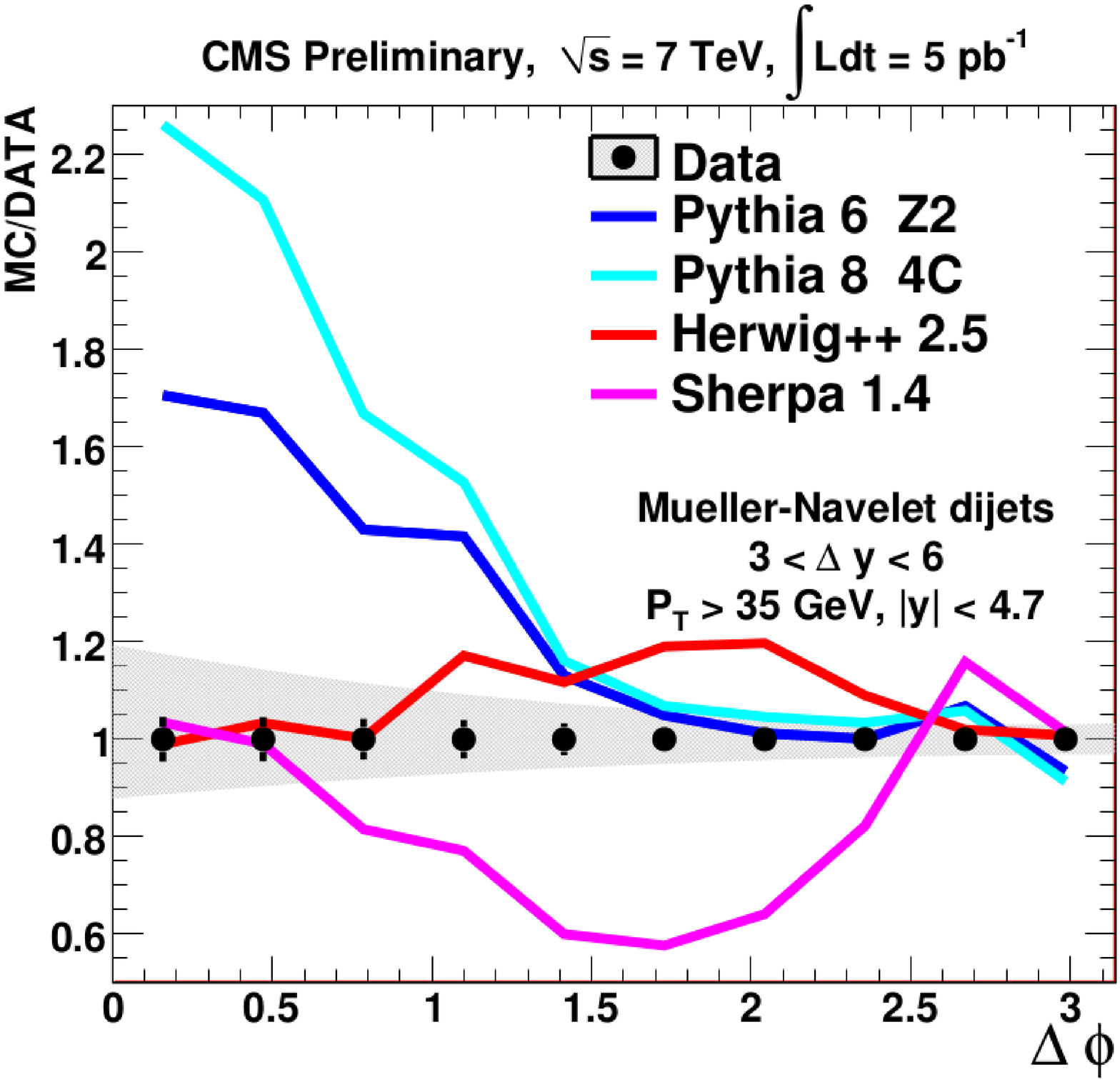}

\includegraphics[width=0.45\textwidth]{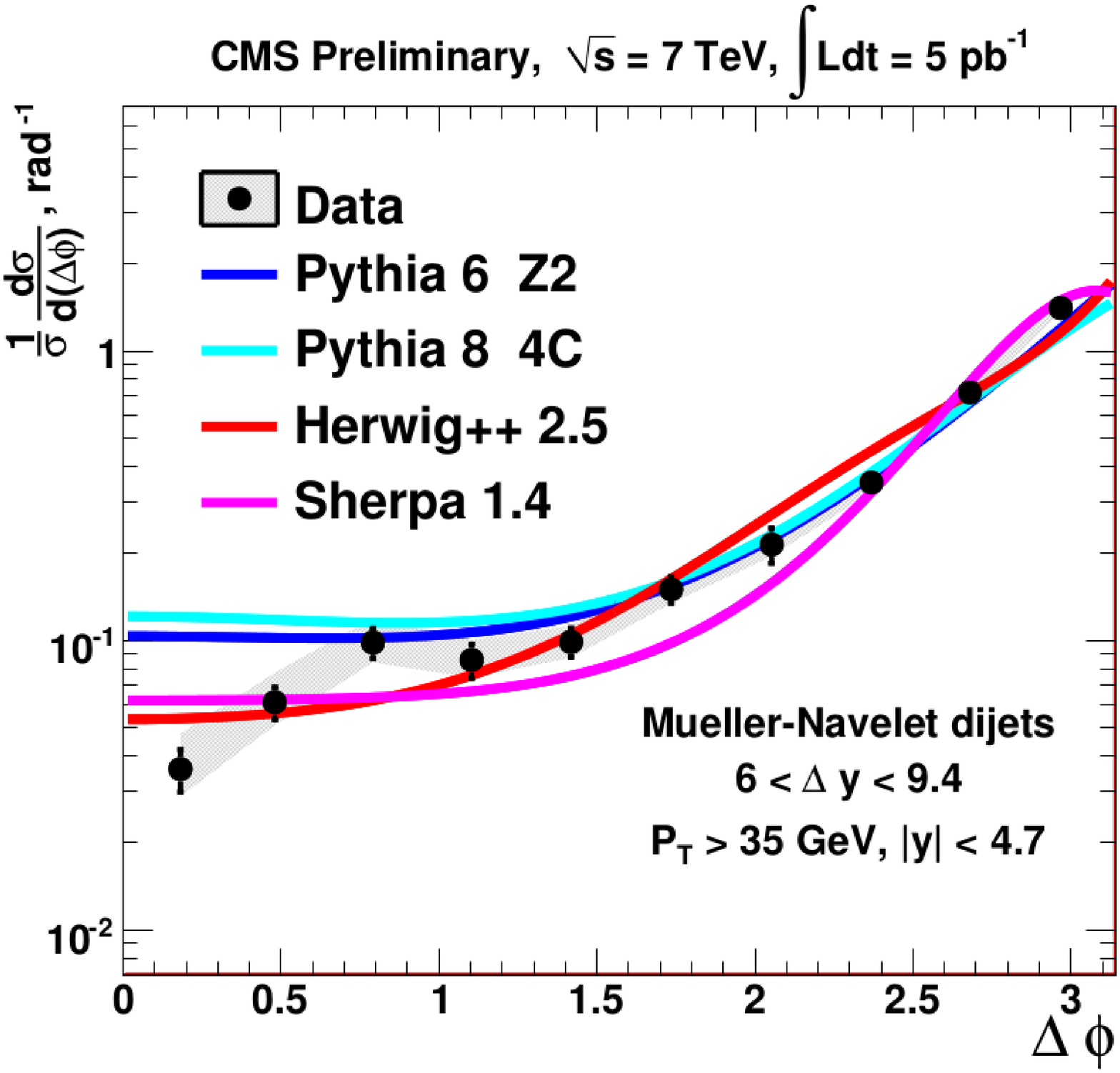}
\includegraphics[width=0.45\textwidth]{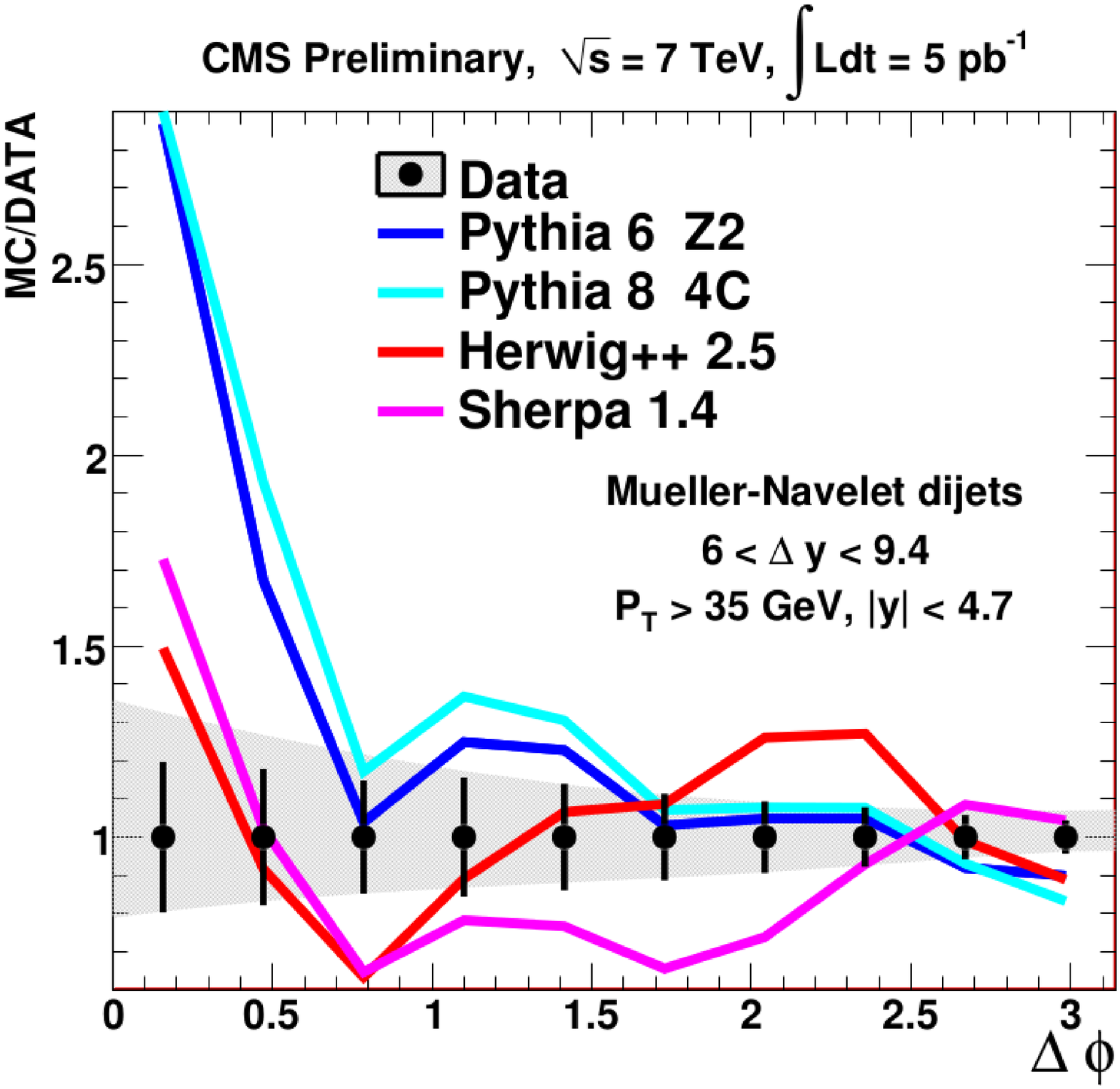}

\caption{Azimuthal--angle decorrelations of jets widely separated in rapidity compared with different Monte Carlo 
predictions~\cite{mn1}.}
\label{fig:azimuthal_decorrelations1}
\end{center}
\end{figure}

\begin{figure}[htp]
\begin{center}
\includegraphics[width=0.32\textwidth]{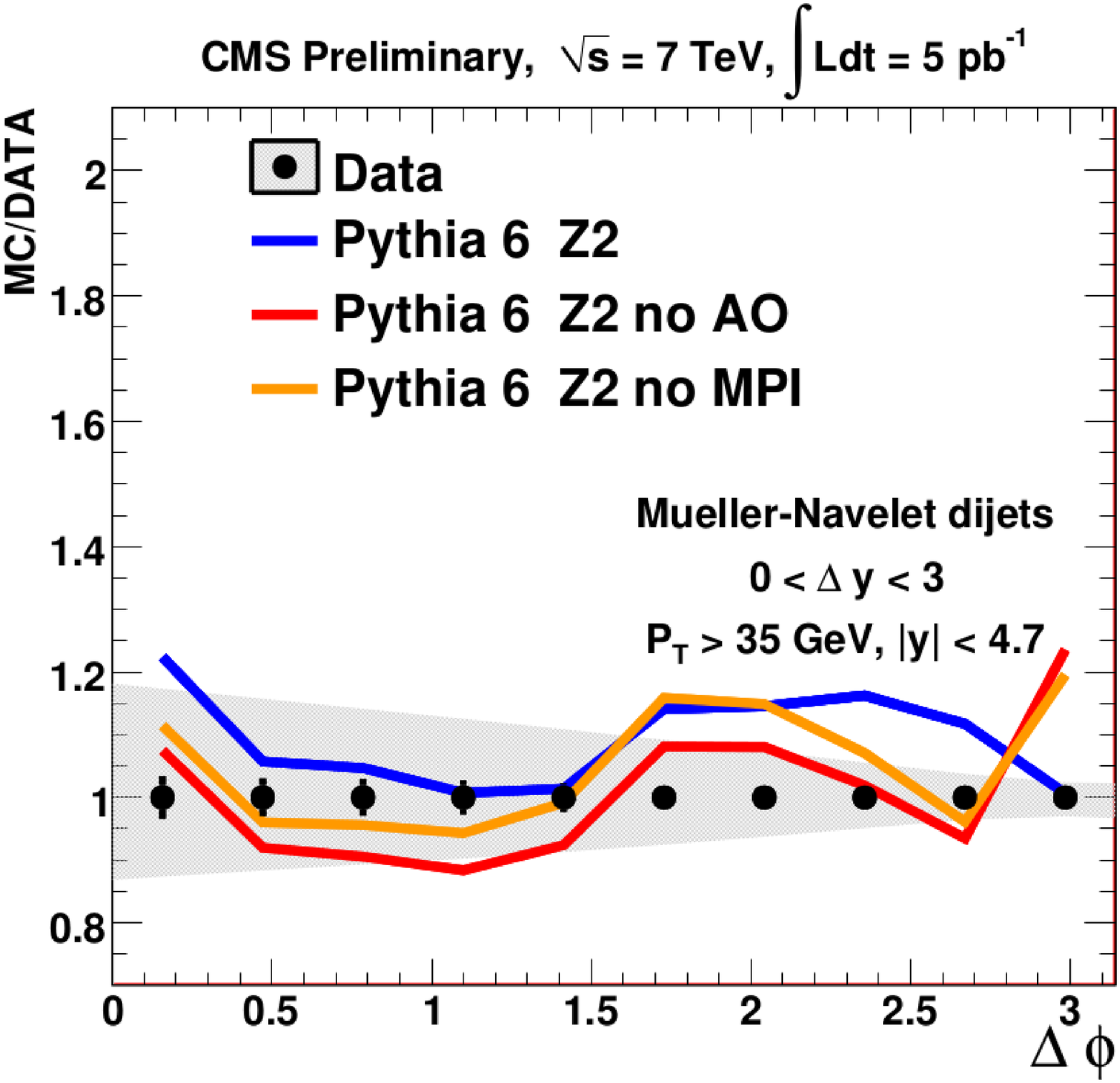}
\includegraphics[width=0.32\textwidth]{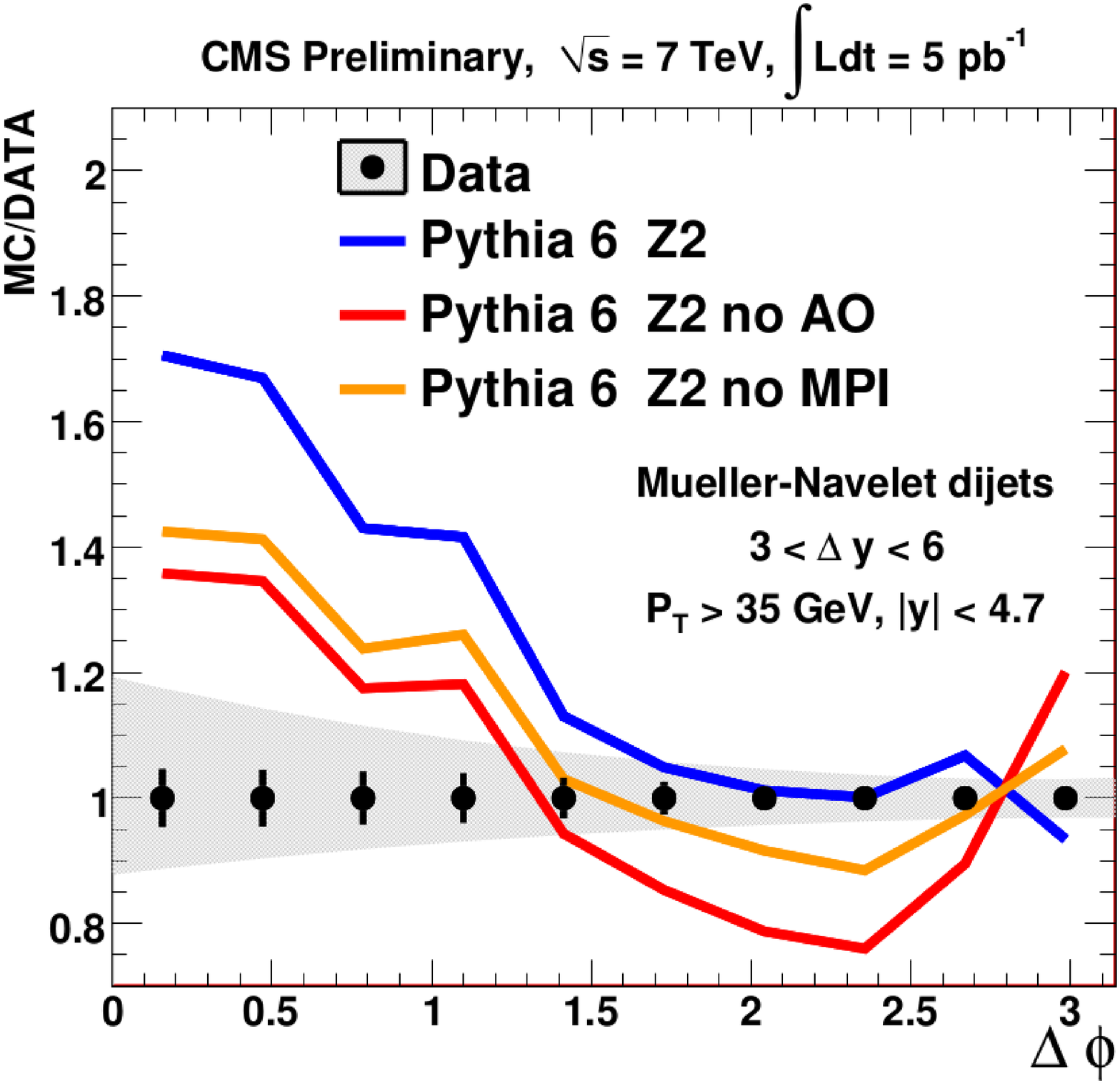}
\includegraphics[width=0.32\textwidth]{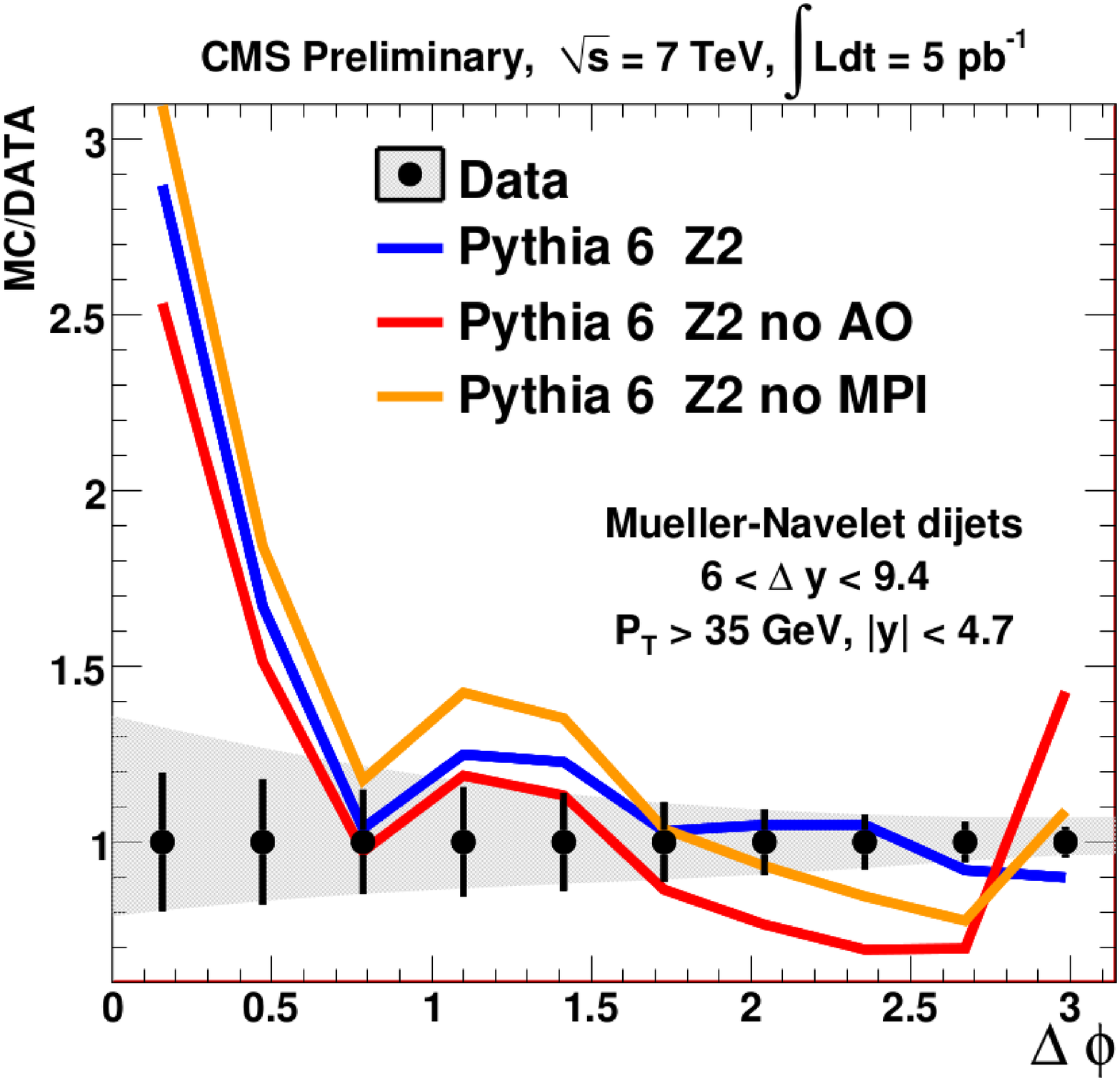}

\caption{Azimuthal angle decorrelations of jets widely separated in rapidity compared with different PYTHIA6 predictions~\cite{mn1}.}
\label{fig:azimuthal_decorrelations2}
\end{center}
\end{figure}

The reconstruction and correction procedure is similar as for the inclusive forward jet production~\cite{mn1}. 
Mueller-Navelet jets are the dijet pair with the highest rapidity separation. In this analysis only jets with $p_T$ above 35 GeV
and $|\eta| < $ 4.7 were considered. The azimuthal angle decorrelations of jets widely separated in rapidity is presented in 
figures \ref{fig:azimuthal_decorrelations1} and \ref{fig:azimuthal_decorrelations2} as function of rapidity separation.

The first row of figure \ref{fig:azimuthal_decorrelations1} displays the azimuthal angle difference $\Delta\phi$ for jets with a 
rapidity separation $\Delta y$ less than 3. {\sc Pythia 6} and {\sc Herwig ++} describe the data within uncertainties,
while {\sc Pythia 8} and {\sc Sherpa 1.4}~\cite{sherpa} with parton matrix elements matched show deviations at small and intermediate $\Delta\phi$. 
The second row shows $\Delta\phi$ for a rapidity separation between 3 and 6. {\sc Herwig ++} provides the best description,
but all predictions show deviation beyond the experimental uncertainties. The last row shows 
the azimuthal--angle difference for $\Delta y$ between 6 and 9. The dijets are strongly decorrelated. {\sc Herwig ++}
provides the best description while {\sc Pythia 6} and {\sc Pythia 8} fail for the lower $\Delta\phi$ region.

The figure \ref{fig:azimuthal_decorrelations2} shows $\Delta\phi$ for Mueller-Navelet jets with different 
rapidity separations compared with with different {\sc Pythia 6} predictions. The contributions of the angular 
ordering (AO) and multi--parton interactions (MPI) are very similar. The intermediate $\Delta y$ region
is better described without MPI. Overall the data is better described with AO and MPI.

\section{Fourier coefficients ratio of the average azimuthal cosines}

\begin{figure}[htp]
\begin{center}
\includegraphics[width=0.45\textwidth]{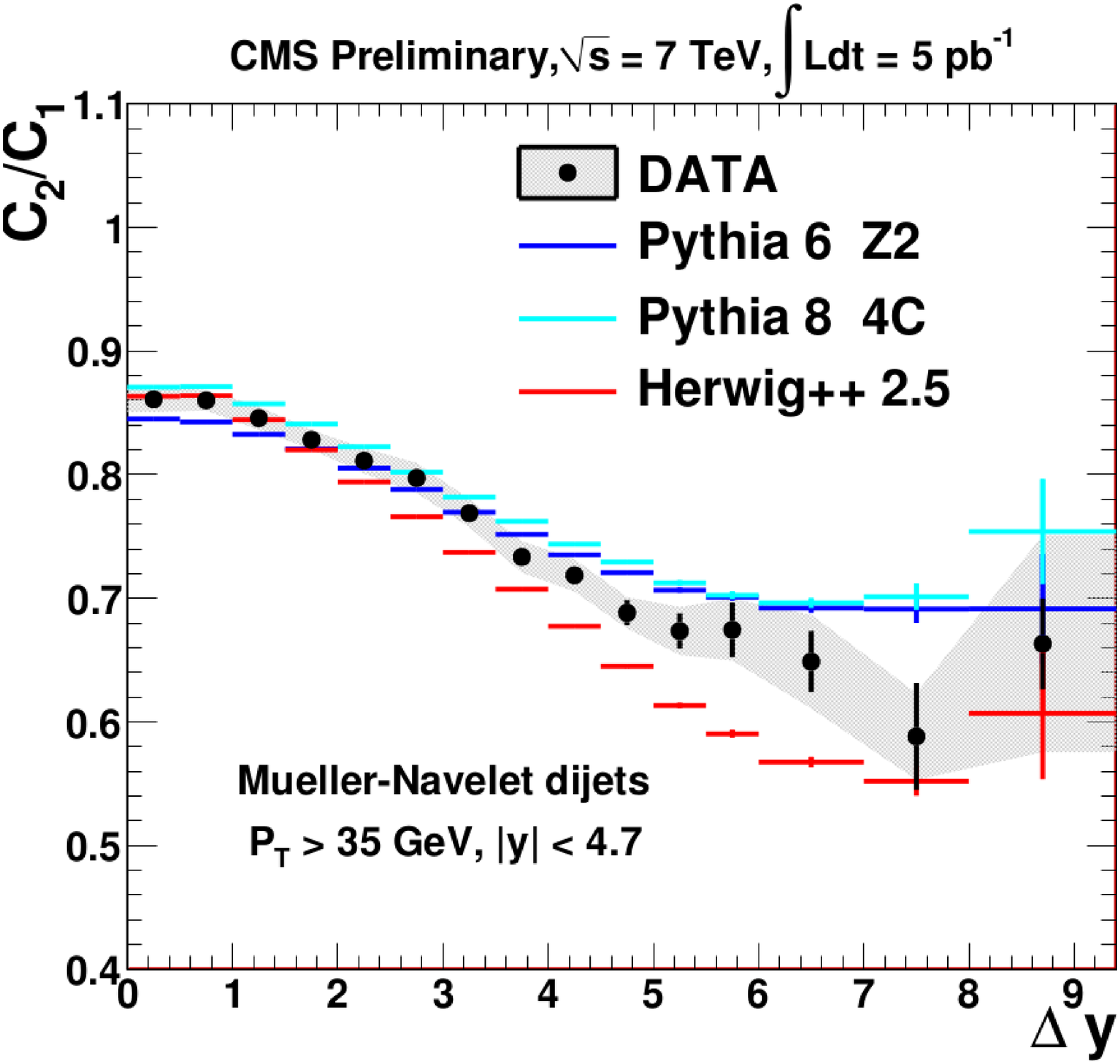}
\includegraphics[width=0.45\textwidth]{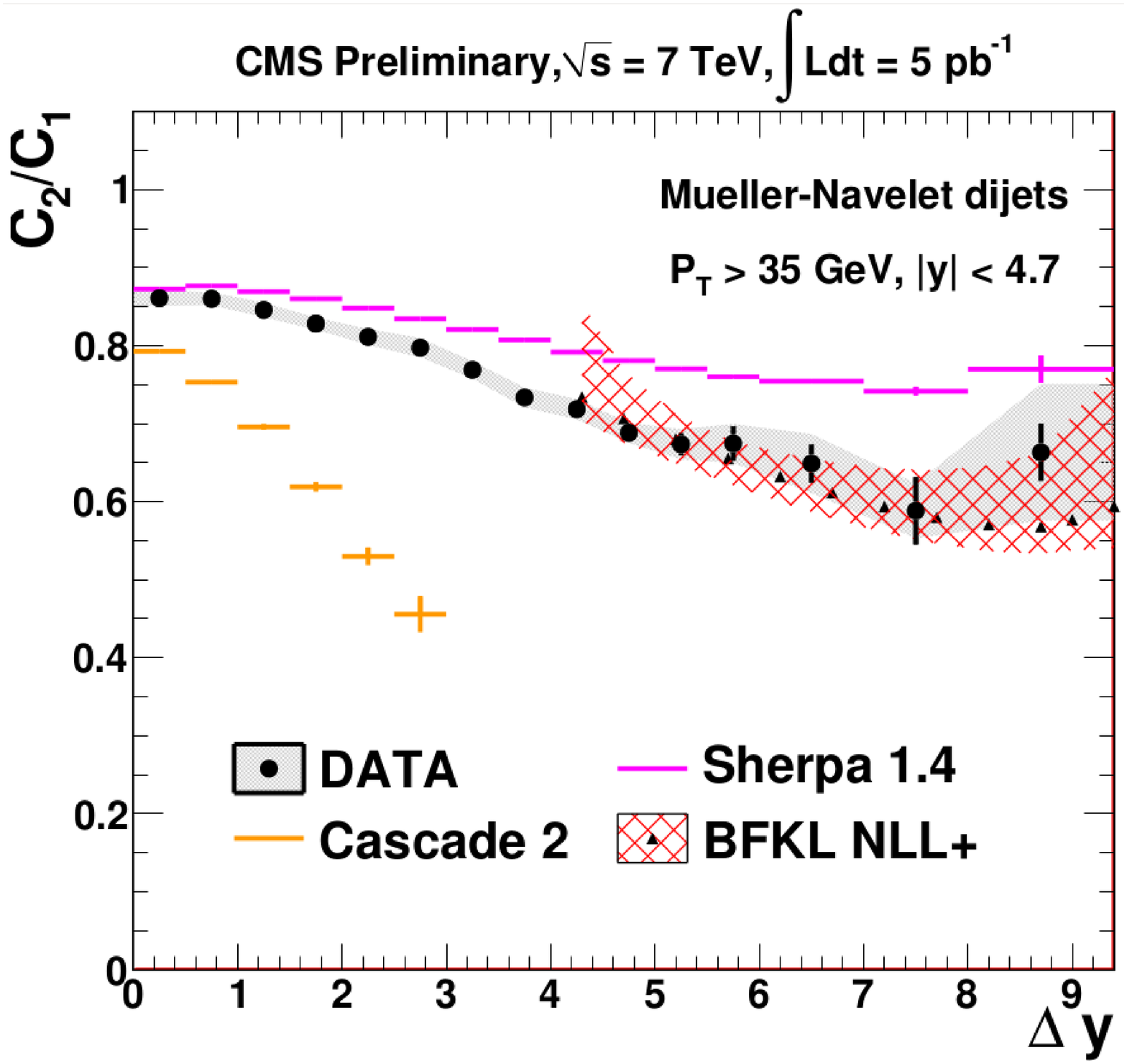}

\caption{Fourier coefficients ratio of the average azimuthal cosines compared with different Monte Carlo predictions~\cite{mn1}.}
\label{fig:average_cosines}
\end{center}
\end{figure}

Using the same selection as in the previous section, the Fourier coefficients of the average cosines 
have been measured~\cite{mn1} and is presented in the figure \ref{fig:average_cosines}.

\begin{equation}
  C_{n} : d\sigma/d(\Delta\phi) \sim \sum C_{n};
  \hspace{5mm}
  C_{n} = <cos(n(\pi - \Delta\phi))>
\end{equation}

DGLAP contributions are expected to partly cancel in the $C_{n+1} / C_n$ ratio, which are 
described the by LL DGLAP--based generators towards low $\Delta y$. {\sc Sherpa}, 
{\sc Pythia 8} and {\sc Pythia 6} overestimate $C_2 / C_1$ while {\sc Herwig} 
underestimate it. The CCFM--based {\sc Cascade} predicts too small $C_{n+1} / C_n$. At $\Delta y > 4$, a
BFKL NLL calculation describe $C_2 / C_1$ within uncertainties.

\section{Ratios of dijets production}

\begin{figure}[htb]
\begin{center}
\includegraphics[width=0.45\textwidth]{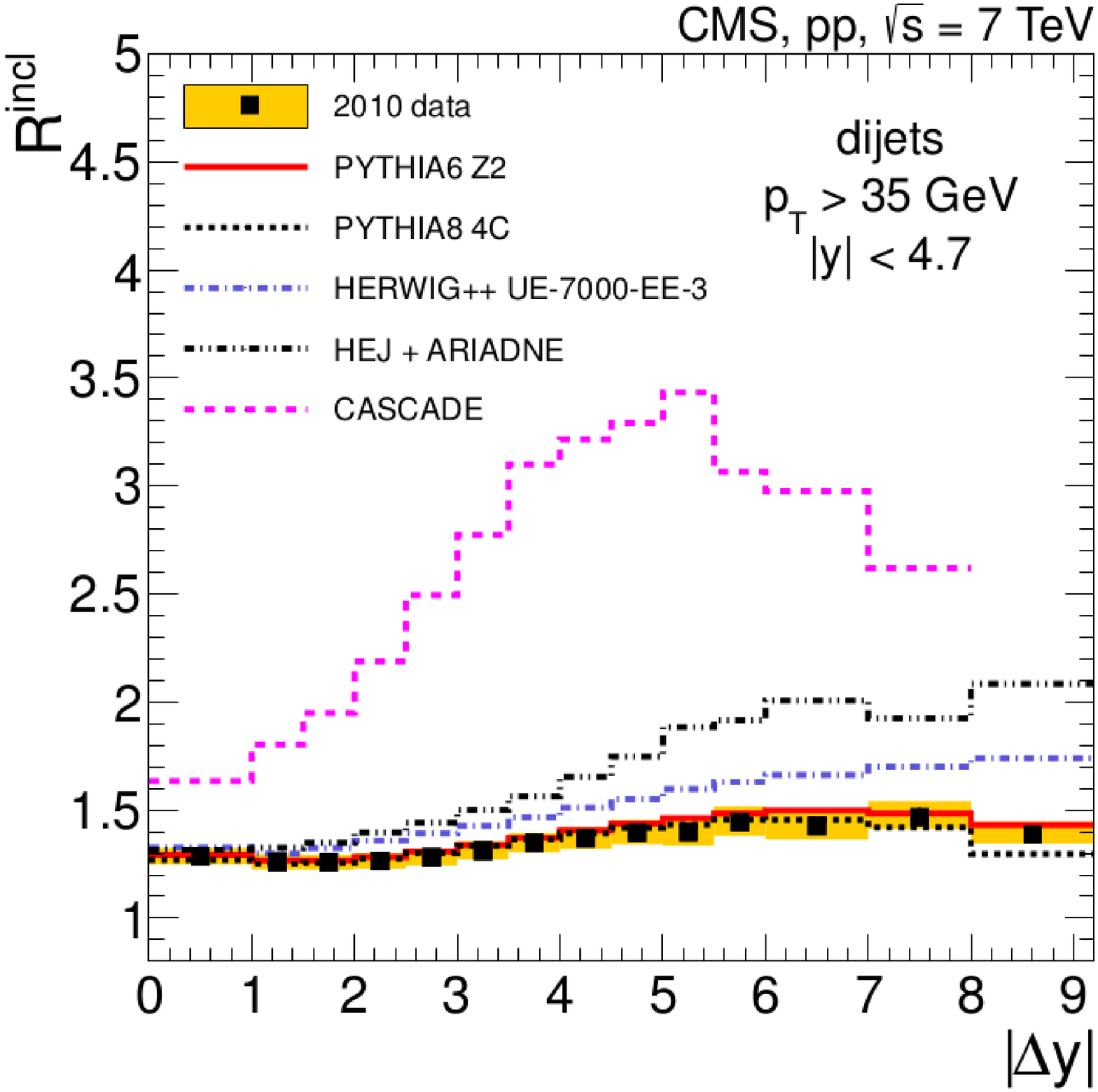}
\includegraphics[width=0.45\textwidth]{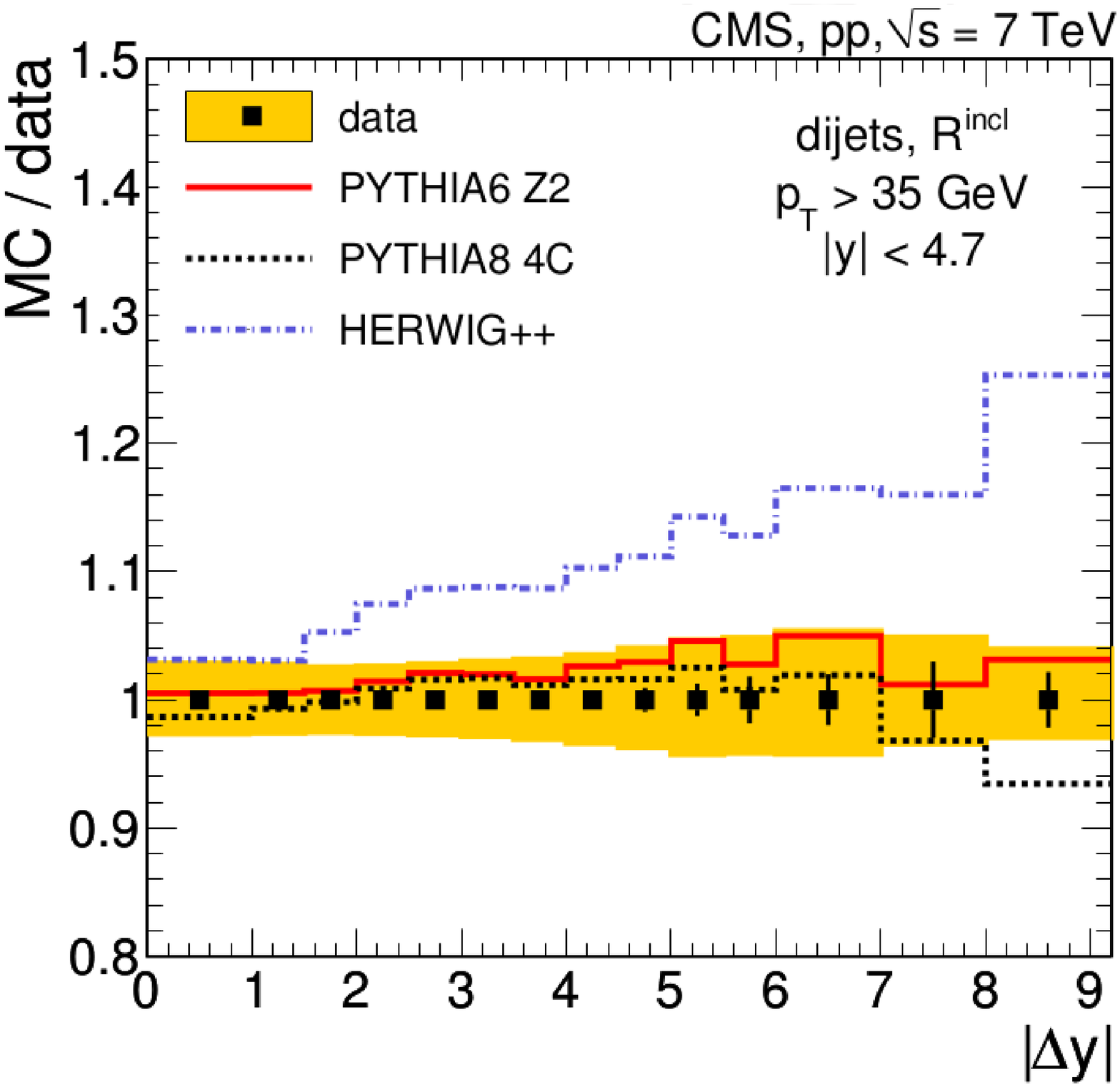}

\caption{Ratios of inclusive/exclusive dijets production compared with different Monte Carlo predictions \cite{kfactor}.}
\label{fig:ratio_dijets1}
\end{center}
\end{figure}

Using jets with $p_T > $ 35 GeV and $|\eta| < $ 4.7 the ratio of the inclusive to exclusive dijet production was measured 
as a function of $\Delta y$ \cite{kfactor}. With increasing $\Delta y$ a larger phase--space for 
radiation is opened. The inclusive dijet sample consists of events with at least 2 jets over the threshold and 
exclusive requires exactly two jets. The ratio of inclusive to exclusive dijet production is 
shown in the figure \ref{fig:ratio_dijets1}. {\sc Pythia 6} and {\sc Pythia 8} agree well with the data while {\sc Herwig ++} 
and {\sc Hej + Ariadne}~\cite{ariadne} overestimate the data at higher $\Delta y$. {\sc Cascade} is completly off. MPI gives only a small contribution.

\begin{figure}[htb]
\begin{center}
\includegraphics[width=0.45\textwidth]{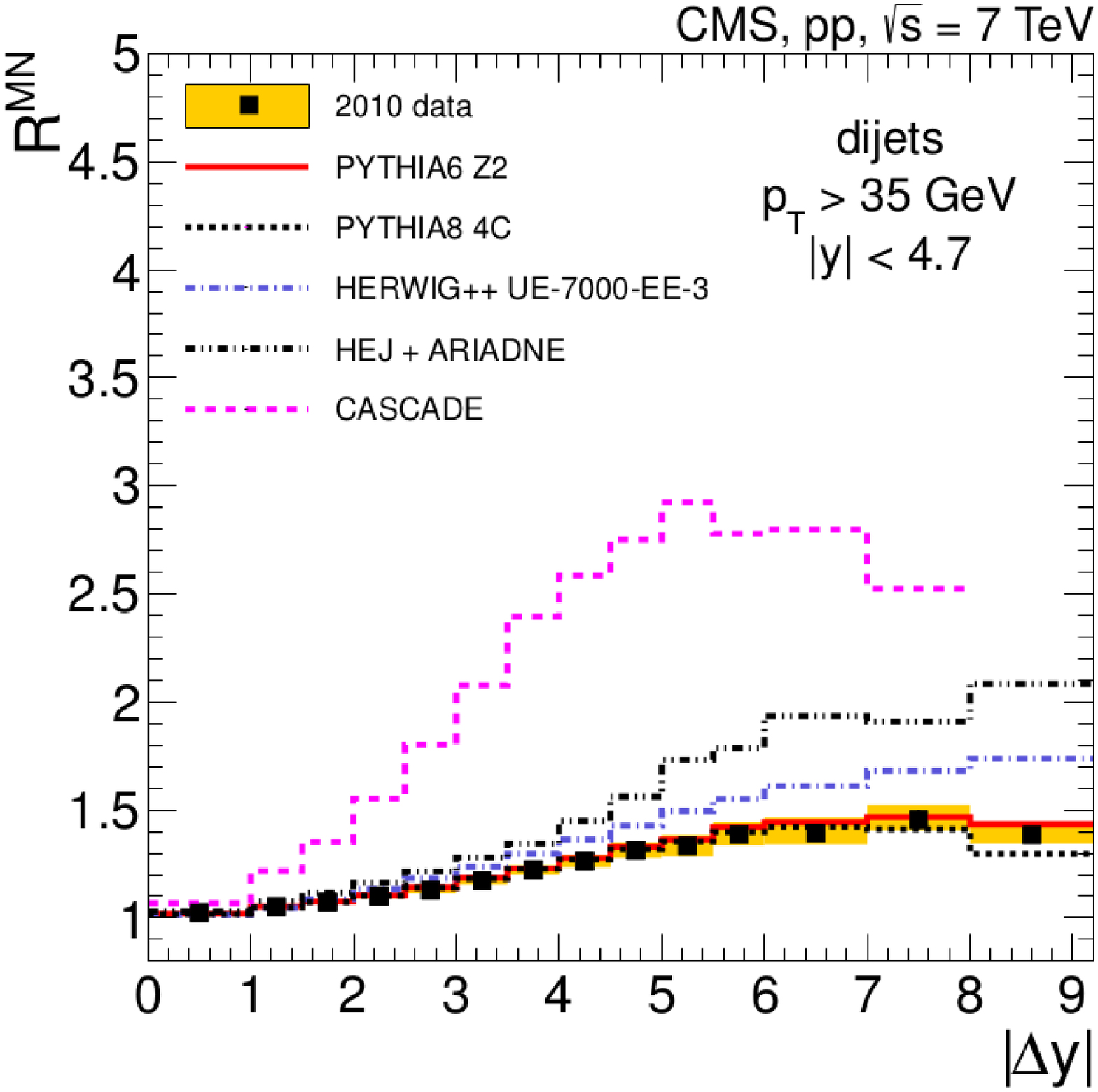}
\includegraphics[width=0.45\textwidth]{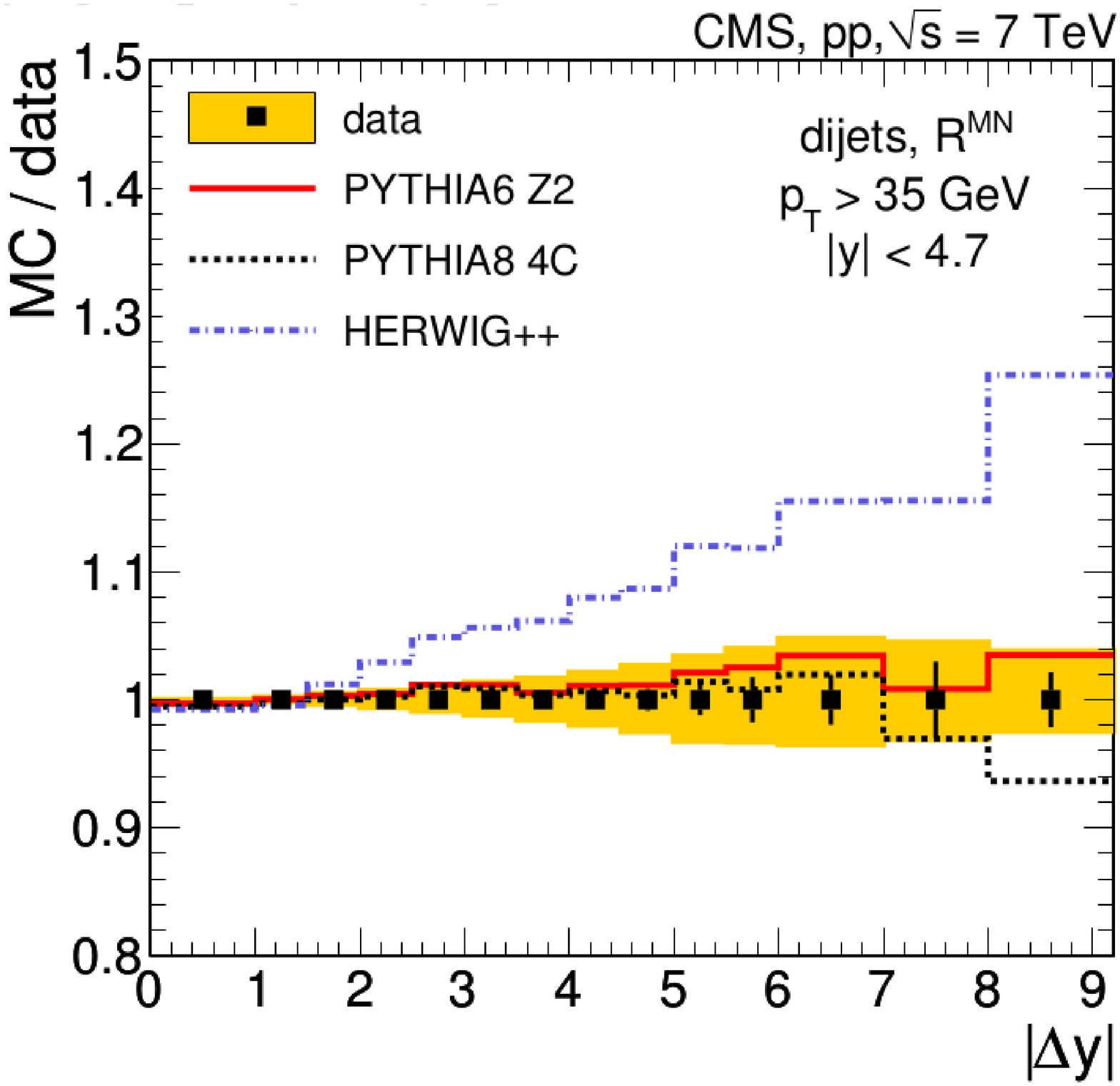}

\caption{Ratios of Mueller-Navelet/exclusive dijets production compared with different Monte Carlo predictions \cite{kfactor}.}
\label{fig:ratio_dijets2}
\end{center}
\end{figure}

The ratio of inclusive to exclusive Mueller-Navelet dijets is 
presented in \ref{fig:ratio_dijets2}. At low $\Delta y$ the ratio of Muller-Navelet over exclusive is, by
definition, smaller than inclusive over exclusive and at higher $\Delta y$ it is the same. The conclusions of the
comparison between data and MC are the same as for the ratio inclusive over exclusive.

\section{Summary}

Inclusive measurements of forward and central--forward jets, are 
reasonably well described by the MC predictions while more exclusive measurements 
are poorly described. A summary of the MC description is presented in table \ref{summary}.
The DGLAP--based generators, {\sc Pythia} and {\sc Herwig}, seem to do a better job than the BFKL--inspired {\sc Cascade}. The 
effort of description of the underlying events, development of the parton showers and tuning of {\sc Pythia} 
and {\sc Herwig} play an huge role into this result.

\begin{table}[htb]
\begin{tabular}{|l|c|c|c|c|c|}
\hline
Observable & {\sc Pythia} & {\sc Herwig} & {\sc Cascade} & {\sc HEJ} \\
\hline
Forward jet $p_{T}$ & Good & Acceptable & Acceptable & Good \\
\hline
Central-forward jet $p_{T}$ & Bad & Acceptable & Bad & Good \\
\hline
Azimuthal correlations & Acceptable & Good & Bad & -- \\
\hline
Fourier coefficients ratio & Acceptable & Bad & Bad & -- \\
\hline
Dijet ratios & Good & Acceptable &  Bad & Bad \\
\hline
\end{tabular}
\caption{Monte Carlo description of the measurements}
\label{summary}
\end{table}

\section*{Acknowlegements}

To the CMS collaboration for the oportunity to join this conference and to Hannes Jung 
for supervision in writing this proceeding.

\bibliographystyle{apsrev4-1}

\end{document}